\documentstyle[aps,prl,multicol,epsf]{revtex}
\tighten
\begin{document}

%
%
\title{How large can be SDW and CDW amplitudes in underdoped 
cuprates}
\author{M. Eremin$^1$, S. Varlamov$^{1,2}$, and I. Eremin$^1$}
\address{$^1$Kazan State University, 420008 Kazan, Russian Federation ,\\
$^2$Cottbus Technical University, 03013 Cottbus, Germany}
\date{\today}
\maketitle
%
%
\begin{abstract}
Self-consistent calculation of spin (charge) density wave order (SDW/CDW) 
parameters have been performed for bilayered cuprates on the basis of a
singlet correlated band model. Evolution of the Fermi surface in the strongly underdoped
regime is described by using a two-band approach. The smooth development of the pseudogap 
formation temperature is explained from underdoped to overdoped states and
the Fourier amplitudes $<s_q>$ (spin) and $<e_q>$ (charge) modulations 
have been calculated. We have 
found a maximum of the incommensurability for doping $0.09 \div 0.11$
holes per copper site.
\end{abstract}
%
%
\pacs{PACS:74.20.-z, 74.72.-h,71.27.+a}
%
%

\begin{multicols}{2}

\section{INTRODUCTION}
Our calculations were motivated in part by NMR experiments searching the
spatial inhomogeneity of charge and spin density distributions in underdoped
cuprates \cite{Haase}. These investigations are connected with the very important problem of
high-temperature superconductivity because the nature of a pseudogap is not completely
understood. Therefore we continue our previous examination of the singlet correlated band
model for layered cuprates \cite{Eremin1} with respect to its ability to describe the observed
doping dependence of a pseudogap formation temperature \cite{Shen,Ding} and Fermi 
surface evolution\cite{LaRosa}.

\section { Basic relations for one band approach}

We start from the Hamiltonian:

\begin{eqnarray}
H=\sum t_{ij}\Psi _i^{pd,\sigma }\Psi _j^{\sigma ,dp}+\sum 
J_{ij}[2(s_is_j)-\frac{n_in_j}2]+
\nonumber \\
+\sum g_{ij}\delta _i\delta _j + H_{CDW} 
\end{eqnarray}
where $\Psi_{i}^{pd,\sigma}$, $\Psi_{j}^{\sigma,pd}$ are quasiparticle 
Hubbard-like operators for the copper-oxygen singlet band, $J_{ij}$ is the
superexchange constant of the copper spin coupling and $g_{ij}$ is 
a screened Coulomb repulsion of the doped holes, 
$1+\delta_i=\sum_{\sigma}\Psi _i^{\sigma,\sigma}+2\Psi _i^{pd,pd}$. 
The quasiparticle interaction $H_{CDW}$ mediated by the phonon field leads to
a CDW transition \cite{Eremin2}. 

In addition to the usuall mean field approach
we have taken into account that anticommutators of
Hubbard-like operators can be affected by the doping index
level per 
one unit cell, spin magnetization and charge modulation (non-Fermi
statistics effect 
(NFS)):

\begin{equation}
P_i^{\sigma }=\Psi _i^{pd,pd}+\Psi _i^{\sigma ,\sigma 
}=\frac{1+\delta_i}2+(-1)^{\frac{1}2-\sigma}s_i^z 
\end{equation}

The appearence of SDW we describe via the Fourier component:

\begin{equation}
s_{q_s}^z=\frac 1N\sum s_i^z\exp (iq_sR_i), 
\end{equation}

where $q_s$ is the instability wave vector with respect to SDW formation. 
CDW can be taken into account in the same wayr by introducing 
the Fourier component of doped holes:

\begin{equation}
e_{q_e}=\frac 1N\sum \delta _i\exp (iqR_i) 
\end{equation}

where $q_e$ is the instability wave vector with respect to CDW formation.
It is easy to show that the homogeneous part ($\delta _0$) of the doped hole 
function per one unit cell ($\delta _i$)
does not contribute to the expectation value $e_q$ $(i.e.\delta _i=\delta _0+e_i)$. 
Generally the wave vectors of CDW $(q_e)$ and SDW $(q_s)$ can be different. 
Below we consider both as commensurate
wave vectors $q_s=q_e=(\pi,\pi)$ and incommensurate wave vectors 
$q_s=q_e=(\pi\pm\varepsilon_x,\pi\pm\varepsilon_y )$.

Using Roth's decoupling procedure \cite{Roth} in the framework of a 
linear approximation we can write:

\begin{eqnarray}
\frac{<P_i^{\sigma }P_j^{\sigma }+
\Psi _i^{\sigma ,\bar{\sigma}}
\Psi _j^{\bar{\sigma} ,\sigma }>}{<P_j^{\bar{\sigma} }>} 
=\frac{1+\delta _0}2[1+\frac{4<s_is_j>}{(1+\delta _0)^2}]- 
\nonumber \\
-(<\frac{e_i}2>+(-1)^{\frac{1}2-\sigma}<s_i^z>)
[1-\frac{4<s_is_j>}{(1+\delta _0)^2}]
\end{eqnarray}

where the angular brackets correspond to the thermodynamic expectation values. Then
the equations of motion are written as:

\begin{eqnarray}
i\hbar \frac \partial {\partial t}\Psi _k^{\sigma ,pd}&=&\varepsilon
_k\Psi _k^{\sigma ,pd}+\eta _{k+q}^{\sigma }\Psi
_{k+q}^{\sigma ,pd} 
\nonumber \\  
i\hbar \frac \partial {\partial t}\Psi _{k+q}^{\sigma ,pd}&=&\varepsilon 
_{k+q}\Psi_{k+q}^{\sigma ,pd}+\eta _k^{\sigma }\Psi _k^{\sigma ,pd}
\end{eqnarray}

where

\begin{eqnarray}
\eta _{k+q}^{\sigma }&=&[t_{k+q}-\frac 4{(1+\delta 
_0)^2}<s_is_j>t_k]\nonumber \\
&& {} <\frac{e_q}2-(-1)^{\frac{1}2-\sigma}s_q^z> + G_{k+q}^{\sigma } + G_{k}^{ph}
\end{eqnarray}

and $\varepsilon_k$ is the energy dispersion in the normal state.
The CDW order parameter is determined by the relations \cite{Eremin3}:

\begin{eqnarray}
G_{k+q}^{\sigma }=-\frac 2{(1+\delta _0)N}\sum_{k'} \{j_{k'-k}<\Psi
_{k'+q}^{pd,\bar{\sigma} }\Psi _{k'}^{\bar{\sigma} ,pd}>+
\nonumber\\
+g_{k'-k}<\Psi
_{k'+q}^{pd,\sigma }\Psi _{k'}^{\sigma ,pd}>\},
\end{eqnarray}

\begin{eqnarray} 
G_{k}^{ph}&=&\sum_{\omega_{q}} [ A(\omega_{q})-B(\omega_{q})\times
\nonumber\\
&& {} \times\frac{\left( \hbar\omega_{q}\right)^{2}
\Theta \left(\hbar\omega_{D}-|\varepsilon_{k}|
\right)\Theta\left(\hbar\omega_{D}-|\varepsilon_{k}-
\varepsilon_{k+q}|\right)}{\left(\varepsilon_{k}-
\varepsilon_{k+q}\right)^{2}-\left(\hbar\omega_{q}\right)^{2}}]
\end{eqnarray}

where $\omega_{q}=40 meV$ is the frequency of active phonon mode in the CDW transition. 

The thermodynamic values of the Fourier component $<e_q>$ and $<s_q>$ are calculated 
self-consistently:

\begin{eqnarray}
e_q&=&\frac 12\sum_{k} [<\Psi _{k+q}^{pd,\sigma }\Psi _k^{\sigma 
,pd}>+<\Psi_{k+q}^{pd,\bar{\sigma} }\Psi _k^{\bar{\sigma} ,pd}>] 
\nonumber \\
s_q^z&=&\frac 12\sum_{k} [<\Psi _{k+q}^{pd,\bar{\sigma} }\Psi _k^{\bar{\sigma}
,pd}>-<\Psi _{k+q}^{pd,\sigma }\Psi _k^{\sigma ,pd}>] 
\end{eqnarray}

The correlation function is determined by

\begin{equation}
<\Psi _{k+q}^{pd,\sigma }\Psi _{k}^{\sigma ,pd}>=
P\frac{\eta_{k+q}^{\sigma }}{E_{1k}^{\sigma}-E_{2k}^{\sigma}}
[f(E_{1k}^{\sigma})-f(E_{2k}^{\sigma })]
\end{equation}

with

\begin{equation}
E_{1k,2k}^{\sigma }=\frac{\varepsilon _k+\varepsilon _{k+q}}2\pm 
\frac12[(\varepsilon _k-\varepsilon _{k+q})^2+4\mid 
\eta _{k+q}^{\sigma}\mid ^2]^{1/2} \\
\end{equation}

\section{Energy dispersion at low doping}

The one band approach is usually valid for a large enough doping level. Therefore
if $\delta$ goes to zero we explored the two band model for layered 
cuprates \cite{Eremin1,Plakida,Eremin4}. 
In this case the energy dispersion is:

\begin{equation}
\varepsilon_k=\frac{E_k^{dd}+E_k^{pp}}2+ 
\frac12[(E_k^{dd}-E_k^{pp})^2+4E_k^{dp}E_k^{pd}]^{1/2} \\
\end{equation}

 where 

\begin{equation}
E_k^{dd}=\varepsilon _d+\sum[\frac{1-\delta _0}2+\frac 
2{(1-\delta_0)}<s_is_j>] t_{ij}\exp (ikR_{ij}) \\
\end{equation}

\begin{equation}
E_k^{pp}=\varepsilon _p+\sum[\frac{1+\delta _0}2+\frac 
2{(1+\delta_0)}<s_is_j>] t_{ij}\exp (ikR_{ij})
\end{equation}

are dispersions of the lower Hubbard copper band and the singlet-correlated oxygen
band, respectively and

\begin{eqnarray}
E_k^{dp}=\sum[\frac{1+\delta _0}2-\frac 
2{(1-\delta_0)}<s_is_j>] t_{ij}^{12}\exp (ikR_{ij}) \\
E_k^{pd}=\sum[\frac{1-\delta _0}2-\frac 
2{(1+\delta_0)}<s_is_j>] t_{ij}^{12}\exp (ikR_{ij})
\end{eqnarray}

describe their hybridization. If $\delta _0$  goes to zero the spin dependent
factors in $E_k^{dd}$  and 
$E_k^{pp}$ become zero due to antiferromagnetic correlations. At the same time,
as one can 
see from the expressions for $E_k^{pd}$  and $E_k^{dp}$  the antiferromagnetic correlations
enhance the 
interband coupling. So, at small doping level the energy dispersion will be
determined by the expression

\begin{eqnarray}
\varepsilon_k(\delta &\rightarrow& 0)=\frac{\varepsilon_d+\varepsilon_p}2+
\nonumber\\ 
&& +\frac12\sqrt{(\varepsilon_p-\varepsilon_d)^2+16[t_1^{12}(\cos{k_x}+\cos{k_y})]^2} 
\end{eqnarray}

where $t_1^{12}$ is a transfer integral between nearest neighbor copper sites.
From (18) one can obtain that the bottom of the band 
corresponds to the point $(\frac{\pi}2,\frac{\pi}2)$ of the Brillouin 
zone. This result coincides with experimental data \cite{LaRosa}. 
We stress here that in contrast 
to Chubukov's model of the normal state of lightly doped cuprates \cite{LaRosa} or
slave boson approximations \cite{Ubbens} any long range order 
is unnecessary in our two band approximation. 
The expression (18) is valid at $<s_i^z>=0$, i.e. for the usual
paramagnetic phase.

\section{Numerical results and discussion}

The spectral weight of the singlet band changes  with doping level
$\delta$ (hole concentration per one unit cell of  bilayer) as
$2\delta/\left(1+\delta\right)$. Thus, the condition of 
half-filling (which approximately corresponds to the optimal doping level )
yields $\delta_{opt}=1/3$. Because the bilayer unit cell  
contains two copper sites we conclude that 
$T_c$ has a maximum near hole concentrations
$x\approx1/6$ per one copper site. 

The evolution of the Fermi surface for three doping levels are presented in
Fig.1. With decreasing doping level the Fermi surface shrinks to the pockets
near the point $(\frac{\pi}2,\frac{\pi}2)$ of the Brillouin zone.

\begin{figure}
\centering
\leavevmode
   \epsfxsize=0.33\textwidth\epsfbox{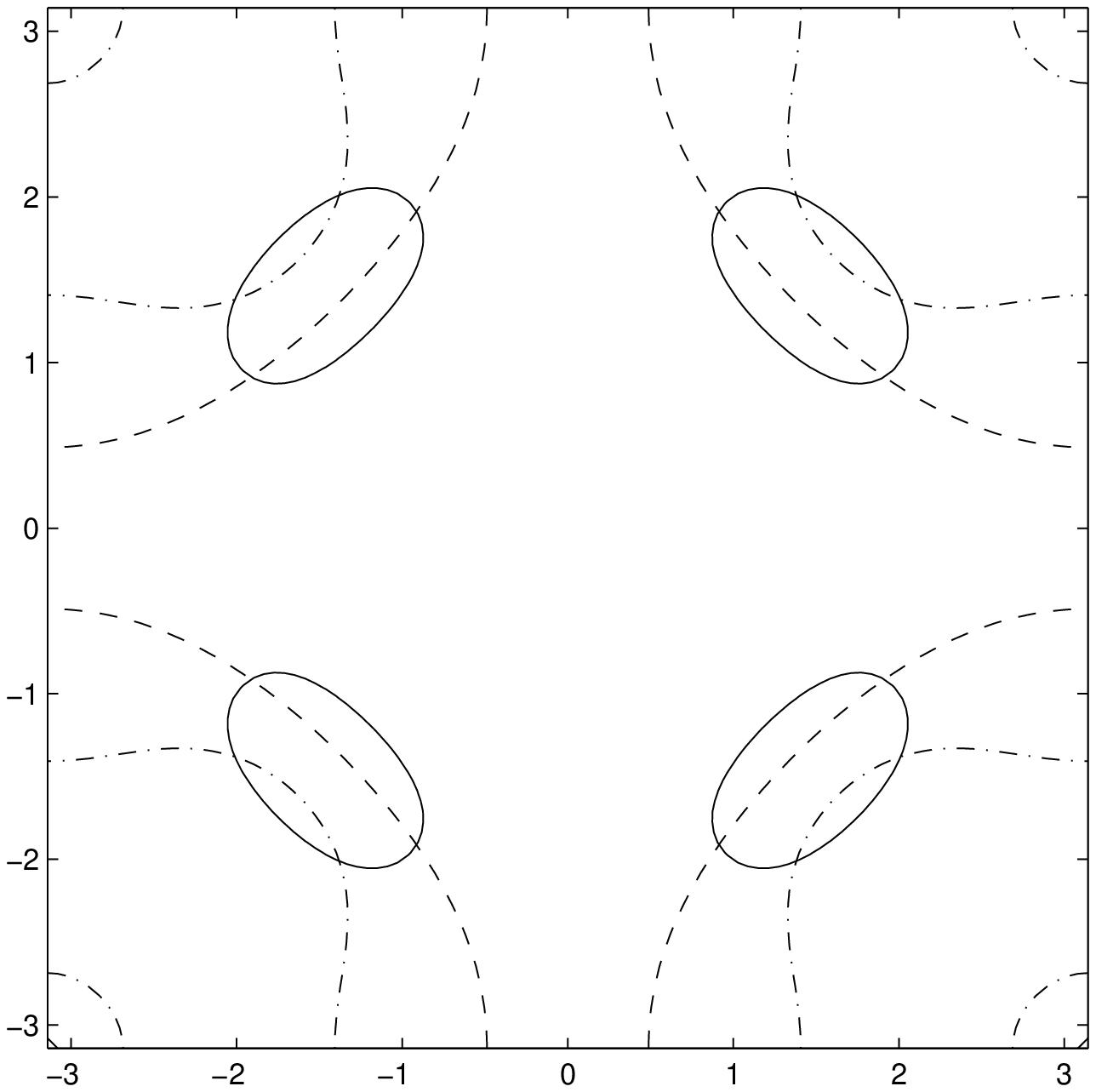}
\put(-80,-9){\makebox(0,0){$k_x$}}
\put(-180,80){\makebox(0,0){$k_y$}}
\vskip0.5cm
\begin{minipage}{0.95\hsize}
\caption{Evolution of the Fermi surface of the singlet band for three doping
levels; $x=0.025$ (solid curve), $x=0.09$ (dash-dotted) and $x=0.17$ ( dashed ).
}\label{fig:1}
\end{minipage}
\end{figure}
\par

This behavior  qualitatively agrees with photoemission 
experimental data \cite{Shen,Ding} and provides a good basis for the analysis of Peierls
like instabilities, which as it was pointed out by many authors
\cite{Eremin2,Markiev,Onufr}, is very sensitive to the topological 
properties of the Fermi surface.

Numerical solutions of the equations (8)-(10) are shown in Fig. 2.

\begin{figure}
\centering
\leavevmode
   \epsfxsize=0.44\textwidth\epsfbox{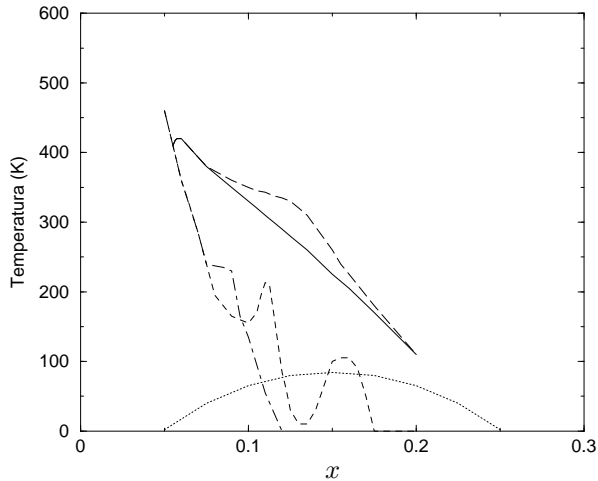}
   \put(-100,-7){\makebox(0,0){$x$}}
\vskip0.5cm
\begin{minipage}{0.95\hsize}
\caption[]{Critical temperatures vs. doping; 
 $T_d^{*}$ (CDW )- long dashed curve,  $T_d^{*}$ (SDW)- 
solid curve, $T_s^{*}$ (CDW)- dashed-dotted, $T_s^{*}$ (SDW)- short dashed. 
The parabola corresponds to the superconducting transition temperature 
$T_c$ (schematically).}\label{fig:2}
\end{minipage}
\end{figure}
\par

The values of hopping integrals in the one band approximation were choosen between first,
second and third neighbors as $t_1=72.5, t_2=0$ and $t_3=12$ (in meV) respectively.
The set of other used parameters is presented in Table 1.
\vspace {1 truecm}

{\small Table 1. The values of spin-spin correlation function and superexchange
plus Coulomb  screening integral at different doping level}

\begin{tabular}{|p{2cm}|p{3cm}|p{3cm}|}
\hline
x & $<S_iS_j>$ & $J_0+G_0$ (meV) \\ \hline
$ 0.20$    & $ -0.0381$   & $ 160$    \\ \hline
$ 0.15$    & $ -0.0728$   & $ 225$     \\ \hline
$ 0.10$    & $ -0.1287$   & $ 320$     \\ \hline
$ 0.05$    & $ -0.2301$   & $ 320$     \\ \hline 
\end{tabular} 
\vspace {1 truecm}

The  short  range interactions (superexchange and screened Coulomb
repulsion), as a rule, yield the d-wave order parameters of CDW and SDW, whereas
NFL-effects, together with phonon mediated interaction, lead to a
anisotropic s-wave pseudogap component.
In general the calculated CDW and SDW order parameters have
$s+id$  symmetry and the temperature dependence of s- and d-components
are different. The critical temperature of the d-component ($T_d^{*}$) is
always higher than $T_s^{*}$ at all considered doping levels. 
Both types of solutions (CDW or SDW) display the correct doping dependence.
In according to photoemission data \cite{Shen,Ding} and NMR \cite{Will} the critical
temperature of the pseudogap goes down when the doping 
decreases. The phase diagram of superconductivity (parabolic line)  is
given schematically as in \cite{Will}. From Eqs. (8), (11) it is clear that the
critical temperature of the d- component CDW ($T_d^{*}$) is not sensitive 
to the external magnetic field. This result agrees well with recent
experimental observation \cite{Gorny}. Calculated mean field critical temperatures  
of CDW ($T_d^{*}$) are  higher than for SDW in the complete doping range.
This result is also consistent with the widely accepted opinion that a transition
towards the so called "stripe" phase in underdoped cuprates
are charge rather than spin  driven \cite{Zachar}.

In Fig.3 we show the magnitude of the instability wave vector 
$Q=(\pi\pm\varepsilon_x,\pi\pm\varepsilon_y )$ vs. doping index.
We have found the maximum of $\varepsilon_y$ around  $x_{CDW} \sim 0.09$ for CDW
and $x_{SDW} \sim 0.11$ for SDW. It would be interesting to check our theoretical
conclusion experimentally.

 It is easy to see from Eq. (10) that for a d-wave order parameter with pure
commensurate wave vector $q=Q=(\pi,\pi)$ both expectation values $<s_q>$ and
$<e_q>$ vanish, but they are finite when the wave vector Q becomes incommensurate.
Because in general the order parameters are complex, the expectation values
$<s_q>$ and $<e_q>$ have real and imaginary components.
Its values at the superconducting transition temperature are presented in Fig.4.

\begin{figure}
\centering
\leavevmode
   \epsfxsize=0.44\textwidth\epsfbox{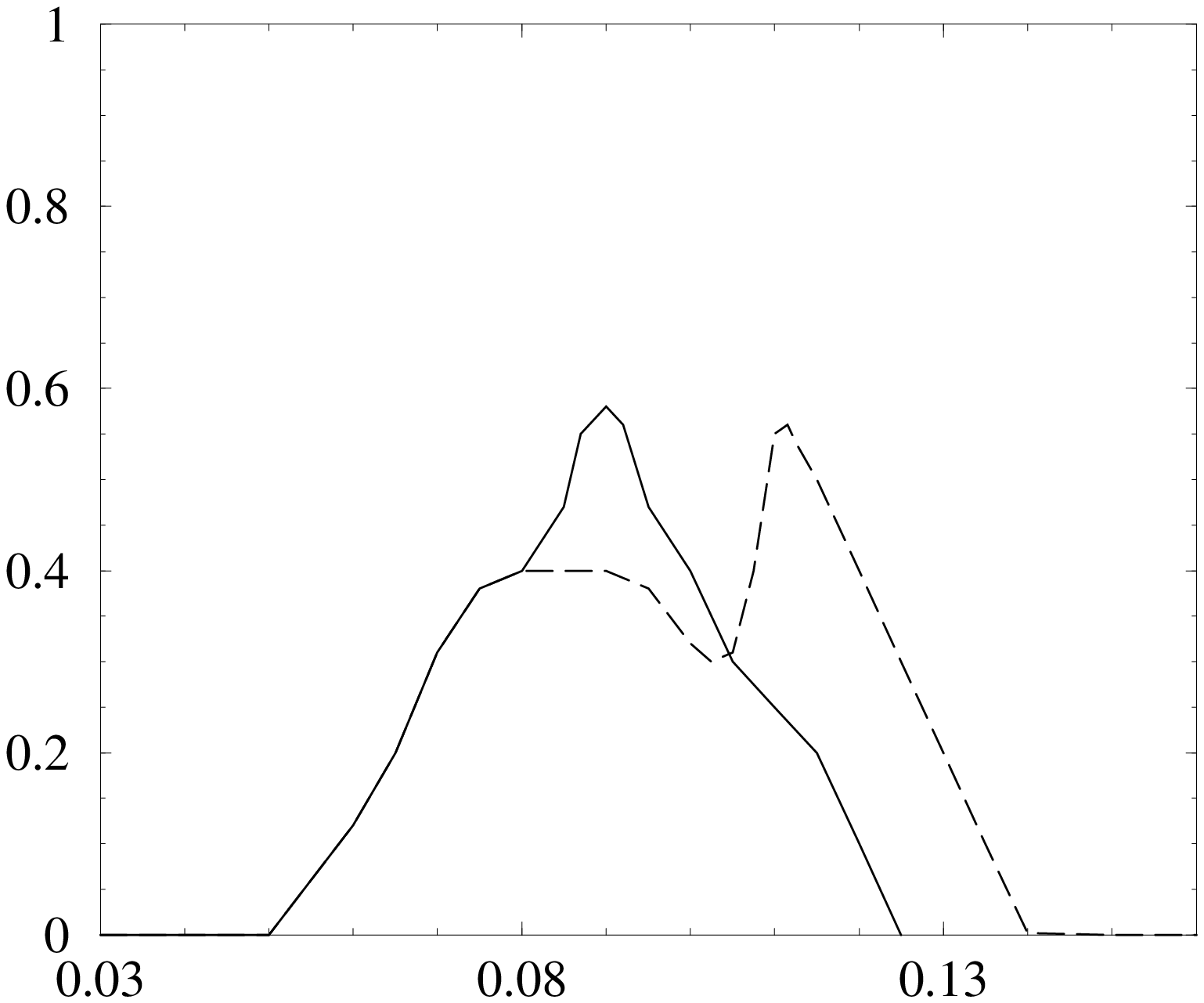}
   \put(-100,-7){\makebox(0,0){$x$}}
 \put(-235,95){\makebox(5,5){$\varepsilon_y$}}
\vskip0.5cm
\begin{minipage}{0.95\hsize}
\caption[]{The variation of incommensurability component $\varepsilon_y$ vs. doping index;
for CDW- solid and for SDW - dashed curve, respectively.}\label{fig:3}
\end{minipage}
\end{figure}
\par

The real component $<e_q>$ is about $0.1-0.15$ at 
temperatures $T \geq T_c$ and disappears  at $T_s^{*}$.

\begin{figure}
\centering
\leavevmode
   \epsfxsize=0.44\textwidth\epsfbox{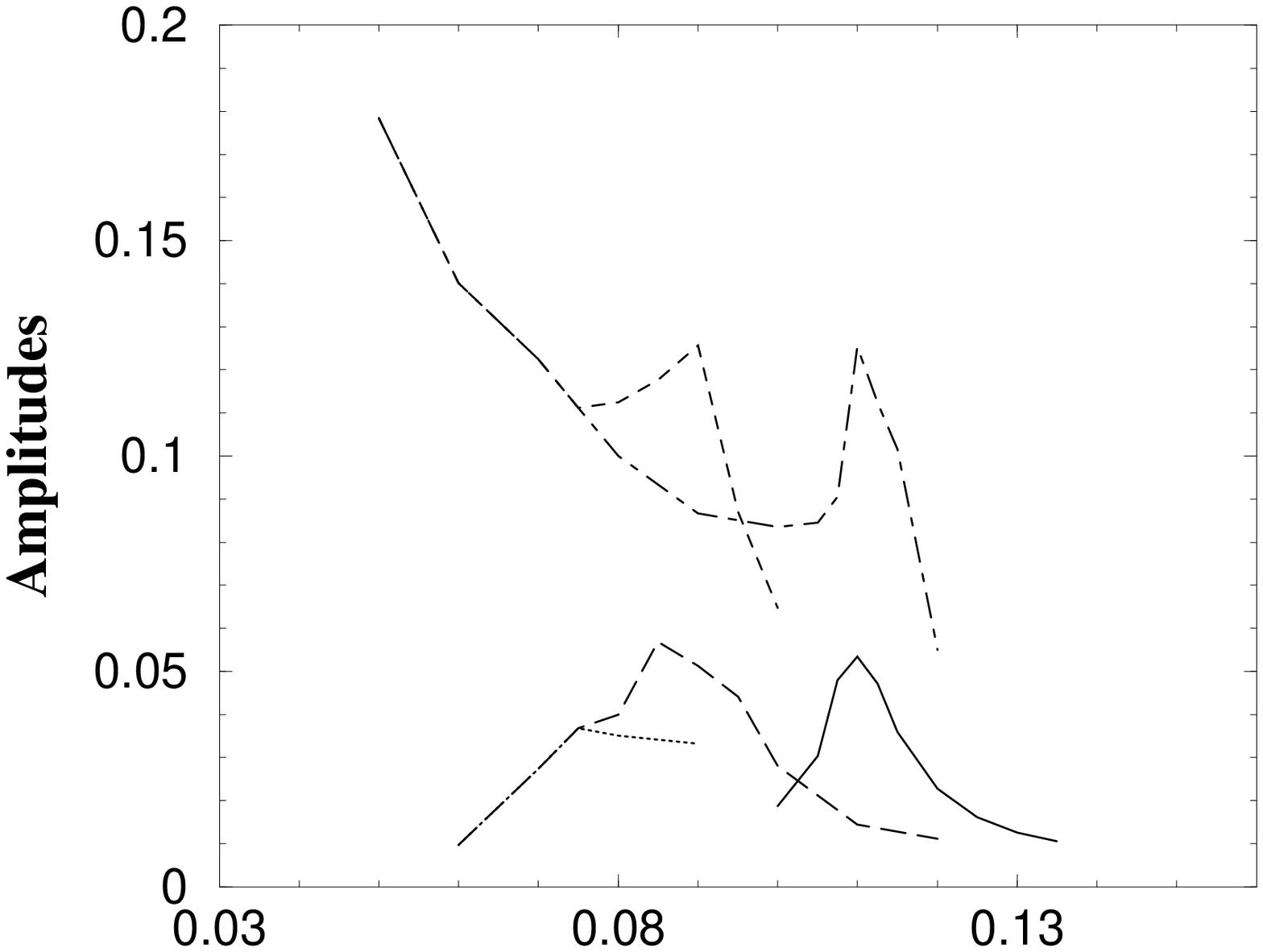}
   \put(-100,-7){\makebox(0,0){$x$}}
\vskip0.5cm
\begin{minipage}{0.95\hsize}
\caption[]{Calculated values $<e_q>$ and $<s_q>$ for temperatures around $T_c$;
$Re<e_q>$(CDW)-dashed curve, $Re<e_q>$ (SDW) - dotted-dashed, $Im<sq>$ (SDW) - solid,
$Im<e_q>$(CDW) -long dashed, $Im<e_q>$(SDW) -dotted. }\label{fig:4}
\end{minipage}
\end{figure}
\par

The imaginary parts of  
$<s_q>$ and $<e_q>$ in general are about  $0.01-0.05$  and stay  up to  $T_d^{*}$ and 
then drastically vanish.

\section{Conclusion}
To conclude we have calculated Fourier amplitudes of CDW and SDW order 
parameters in underdoped cuprates. Our calculation provides an explanation
for the experimentally observed evolution 
of the Fermi surface and the doping dependence of the pseudogap formation 
temperature. The charge instability preforms the spin instability and the 
incommensurability has the maximum at $x \sim 0.1$ holes per one copper site. 
We hope our present calculations will help in better understanding many of
the strange features in the shape and linewidth of NMR in layered cuprates.\\
%
%
{\it Acknowledgments:} 
We would like to thank G.Seibold for helpful comments
and a critical reading of the manuscript.
This work is supported in part by  INTAS, Grant 96-0393.  
One of us (I.Eremin) is grateful for the financial support to 
Russian Ministry of Education (Grant No. 97-0-8.1-7133) 
and to the research program of the International Center for
Fundamental Physics in Moscow (INTAS Grant 96-0457).
%
%

\end{multicols}
\end{document}